\RequirePackage{fixltx2e}
\documentclass{IEEEtran}
\usepackage[T1]{fontenc}
\usepackage{amsmath}
\usepackage{amsthm}
\usepackage{amssymb}
\usepackage{graphicx}

\makeatletter
\theoremstyle{plain}
\newtheorem{thm}{\protect\theoremname}
\theoremstyle{plain}
\newtheorem{lem}[thm]{\protect\lemmaname}
\theoremstyle{remark}
\newtheorem{rem}[thm]{\protect\remarkname}

\usepackage[caption=false,font=footnotesize]{subfig}

\makeatother

\providecommand{\lemmaname}{Lemma}
\providecommand{\remarkname}{Remark}
\providecommand{\theoremname}{Theorem}

\begin{document}

\title{Small Cell Offloading Through Cooperative Communication in Software-Defined
Heterogeneous Networks}

\author{Tao\ Han,\ \IEEEmembership{Member,\ IEEE}, Yujie\ Han, Xiaohu\ Ge,\ \IEEEmembership{Senior\ Member,\ IEEE},
Qiang\ Li,\ \IEEEmembership{Member,\ IEEE}, Jing\ Zhang,\ \IEEEmembership{Member,\ IEEE},
Zhiquan\ Bai,\ \IEEEmembership{Member,\ IEEE}, and Lijun Wang,\ \IEEEmembership{Member,\ IEEE}\thanks{Accepted by IEEE Sensors Journal.The authors would like to acknowledge
the support from the International Science and Technology Cooperation
Program of China under grants 2015DFG12580 and 2014DFA11640, the National
Natural Science Foundation of China (NSFC) under the grants 61471180,
61210002, 61301147, 61301128, 61271224 and 61461136004, the Hubei
Provincial Science and Technology Department under the grant 2013CFB188,
the Hubei Provincial Department of Education Scientific research projects
(No.B2015188), the Fundamental Research Funds for the Central Universities
under the HUST grants 2015XJGH011 and 2015MS038, the grant from Wenhua
College (No.2013Y08), and the Fundamental Research Funds of Shandong
University (No. 2016JC010). This research is partially supported by
the EU FP7-PEOPLE-IRSES, project acronym S2EuNet (grant no. 247083),
project acronym WiNDOW (grant no. 318992) and project acronym CROWN
(grant no. 610524). This research is supported by the National international
Scientific and Technological Cooperation Base of Green Communications
and Networks (No. 2015B01008) and the Hubei International Scientific
and Technological Cooperation Base of Green Broadband Wireless Communications.
\textit{(Corresponding author: Xiaohu Ge.)}}\thanks{Tao Han, Yujie Han, Xiaohu Ge, Qiang Li and Jing Zhang are with the
School of Electronic Information and Communications, Huazhong University
of Science and Technology, Wuhan, 430074 P.R. China, e-mail: \{hantao,
hanyujie, xhge, qli\_patrick, zhangjing\}@hust.edu.cn.}\thanks{Zhiquan Bai is with the School of Information Science and Engineering,
Shandong University, Jinan, 250100 China, e-mail: zqbai@sdu.edu.cn.}\thanks{Lijun Wang is with the Department of Information Science and Technology,
Wenhua College, Wuhan, 430074 P.R. China, e-mail: wanglj22@163.com.}\thanks{Copyright (c) 2016 IEEE. Personal use of this material is permitted.
However, permission to use this material for any other purposes must
be obtained from the IEEE by sending a request to pubs-permissions@ieee.org.}\thanks{Digital Object Identifier 10.1109/JSEN.2016.2581804}}
\maketitle
\begin{abstract}
To meet the ever-growing demand for a higher communicating rate and
better communication quality, more and more small cells are overlaid
under the macro base station (MBS) tier, thus forming the heterogeneous
networks. Small cells can ease the load pressure of MBS but lack of
the guarantee of performance. On the other hand, cooperation draws
more and more attention because of the great potential of small cell
densification. Some technologies matured in wired network can also
be applied to cellular networks, such as Software-defined networking
(SDN). SDN helps simplify the structure of multi-tier networks. And
it's more reasonable for the SDN controller to implement cell coordination.
In this paper, we propose a method to offload users from MBSs through
small cell cooperation in heterogeneous networks. Association probability
is the main indicator of offloading. By using the tools from stochastic
geometry, we then obtain the coverage probabilities when users are
associated with different types of base stations (BSs). All the cell
association and cooperation are conducted by the SDN controller. Then
on this basis, we compare the overall coverage probabilities, achievable
rate and energy efficiency with and without cooperation. Numerical
results show that small cell cooperation can offload more users from
MBS tier. It can also increase the system's coverage performance.
As small cells become denser, cooperation can bring more gains to
the energy efficiency of the network. 
\end{abstract}


\section{Introduction}

\IEEEPARstart{A}{long} with the huge increasing of mobile users,
the current wireless communication system is facing with the great
challenges of system capacity and quality-of-service (QoS) requirements
\cite{Zhang2014CAP}. 5G is expected to achieve gigabit-level throughput
and more varied service capabilities in an energy-efficient way by
2020 \cite{popovski2014ict,ge20145g}. In \cite{Chen2015AIWAC} and
\cite{Chen2015EMC}, mobile cloud and wearable computing is implemented
in 5G to improve Quality of Experience (QoE) and overcome the energy
bottleneck. The deployment of small cells seems to be one of the most
feasible solutions for the areas with a large traffic demand. Heterogeneous
networks (HetNets) where macro and small cells coexist have been widely
applied \cite{damnjanovic2011survey}. BSs in different tiers differ
in transmit power, coverage range, and spatial density. In HetNets,
small cells serve to offload users and traffic data from congested
macro BSs (MBSs) \cite{elsawy2013hetnets}. Compare to traditional
MBSs, small cell base stations (SBSs) have smaller coverage areas
\cite{andrews2012femtocells} but with the advantage of less transmit
power and easy deployment. It will also cut down the construction
cost significantly. \cite{razavi2012urban} shows that there's a significant
gain in power consumption by introducing the small cell tier. However,
multi-tier and denser small cells bring a serious problem, users have
to suffer more severe interference. The deployment of SBSs is still
an effective complement to traditional MBS on coverage and capacity.
Over the multi-tier HetNets, mobile converged networks have become
a focus recently \cite{han2014mobile,Chen2015Cloud}.

Cooperative communication has been deemed as a solution to address
the interference problem \cite{gesbert2010multi,simeone2009local}.
Coordinated multipoint (CoMP) transmission is one of the key technologies
to improve cell edge user data rate and spectral efficiency. Interference
can be exploited or mitigated by cooperation between BSs \cite{irmer2011coordinated}.
\cite{garcia2014coordinated} proposes macrodiversity coordinated
multipoint transmission (MD-CoMP) with user-centric adaptive clustering,
which can significantly improve the coverage performance in different
networks. \cite{baccelli2015stochastic} introduces a novel cooperation
policy which triggers cooperation only when the user lies inside a
planar zone at the edge of the cell. In general, the system's overall
performance can be improved by utilizing cooperation. Considering
the benefits of cooperation among SBSs and small cell offloading in
HetNets, we propose a novel scheme to combine them together, and thus
improving energy efficiency (EE) of the networks. 

The convergence of various technologies is inevitable. Significant
data manipulation, the evolution of equipment and complex protocols
set higher requirements for BS controller. In HetNets, coordination
between different tiers is also difficult because of their different
protocols and interfaces. Software-defined networking (SDN) has shown
its important role in the wired networks by decoupling the control
plane and data plane \cite{nunes2014survey,kreutz2015software}. The
SDN controller is software-based and the entire network is abstracted
in it. This centralization makes it easy to control the network behavior.
Applying SDN to the traditional radio access networks (RAN) is a promising
way. In \cite{arslan2015software} various SDN principles are described
to apply to the RAN. It shows great potential for RAN optimization.
To enable software-defined cellular networks, authors in \cite{li2012toward}
presents several changes and extensions to controller platforms and
BSs. With SDN controller, information exchanged between BSs can be
effectively reduced and backhaul power consumption reduces as well,
thus improving the efficiency of BS cooperation.

\subsection{Related Work}

In recent years, people start to concentrate on cooperation in HetNets.
Most of them put forward the coordination between different tiers,
i.e., the cross-tier cooperation. \cite{gesbert2010multi} presents
an overview of the multi-cell cooperation. It can dramatically improve
the system performance in dense networks where interference emerges
as the key capacity-limiting factor. \cite{nigam2014coordinated}
analyzes the coverage probability of a general user that locates at
an arbitrary location. For the general user, the cooperative set consists
of the BSs with the strongest average received power in each tier.
Similarly, in \cite{jo2012heterogeneous} the authors consider a cell
association based on maximum biased-received-power. It confirms that
a user prefers to connect to a tier with higher BS density and transmit
power. \cite{Ge2015Spatial} proposes a Markov chain based channel
access model and integrates it into random cellular networks. \cite{tanbourgi2014tractable}
and \cite{tanbourgi2014analysis} propose non-coherent joint-transmission
cooperation. The former characterizes the signal to interference and
noise ratio (SINR) distribution for a typical user served by cooperating
BSs. BSs that are sufficiently close are grouped into a cooperative
cluster. \cite{tanbourgi2014analysis} extends the work to heterogeneous
networks. In \cite{lee2012coordinated} the authors demonstrate that
BS cooperation in HetNets achieves higher throughput gains because
of the mitigation of inter-cell interference through cooperation.
Considering the user mobility in real life, \cite{Ge2014Energy} and
\cite{Ge2015User} study the system performance based on the Gauss\textendash Markov
mobile models and individual mobility model respectively. Authors
in \cite{mi2014no} propose a framework called GNV (Global Network
View) in SDN. The related information and states are stored in it
and can be visited by applications and modules. This kind of global
view of the network helps the implement of some complicated technologies
like CoMP. 

In a network with dense small cells, BS density is a considerable
factor that influences the system's performance. To minimize the network
energy consumption, \cite{cao2013optimal} gives the optimal BS density
for both homogeneous and heterogeneous networks. In \cite{quek2011energy}
the authors figure out the BS density ratio that makes the system
get the best energy efficiency. Within the constraints of backhaul
capacity and energy efficiency, authors in \cite{X20155G} prove that
there exists a density limit in 5G ultra-dense networks. Energy efficiency
is what people always endeavor for, especially for the next-gen communications.
Results in \cite{qiao2013base} reveal that BSs' cooperation will
bring gains to the energy efficiency only when most of the BSs participate
in the cooperation. \cite{nie2014energy} formulates a power minimization
problem with the minimum ergodic rate constraint and shows that the
extra deployment of small cells is energy-saving compared to the traditional
macro-only network under its cooperation scheme. Sleeping strategy
is also an effective way to save energy in HetNets \cite{saker2012energy,sun2014energy}. 

One purpose of deploying SBSs is to offload users from the MBS tier.
Small cell's low transmit power makes it offload enough users in an
energy-efficient way. Cell range expansion (CRE) \cite{damnjanovic2011survey}
is a practical technique in which users can be offloaded with biasing.
In \cite{lopez2012expanded} the authors calculate the appropriate
range expansion bias for two different range expansion strategies.
For coverage maximization, the required selection bias is given in
\cite{gupta2014downlink}. \cite{singh2014joint} shows that the offloading
strategy, coupled with resource partitioning is able to improve the
rate of cell edge users in HetNets. \cite{elsawy2013hetnets} evaluates
the load of each network tier and studies different offloading techniques
used to control the load. The traffic offloading is quantified via
the tier association probability. \cite{liu2013energy} proposes two
offloading algorithms, called Traffic Offloading (TO) algorithm based
on the Reference Signal Receive Power and Traffic Offloading based
on Frequency Reuse (TOFFR) algorithm. User association has been studied
to balance the loads among different tiers \cite{son2009dynamic,ye2013user,kim2010alpha}.
With the related load information and centralized manipulation in
SDN, it will be more convenient to address the offloading problems.
In \cite{liu2015distributed}, data offloading optimization is coordinated
by the SDN controller dynamically on the offloading demand and supply.
\cite{arslan2014sdoff} proposes a novel software-defined small cell
offloading control mechanism (SDoff) that can orchestrate the offloading
according to the proposed dissatisfaction parameter and user types.
These applications can also be assisted by software-defined network
function virtualization (VFN) which decouples the network functions
from the hardware \cite{Li2015Software}.

\subsection{Contributions and Organizations}

The main contributions of our work can be summarized as follows:
\begin{itemize}
\item We propose a novel offloading scheme through small cell cooperation
in a software defined 2-tier network where BSs in each tier are distributed
independently according to a Poisson point process (PPP). In the cooperative
model, with the help of the SDN controller, multiple adjacent SBSs
cooperate to transmit data to a specified user if they can jointly
offer stronger signal than the MBS. Thus it can offload more users
from the MBS tier compare to one SBS. For the single BS association,
the user will connect to the nearest BS. We evaluate the association
probabilities to measure the cell's traffic load. 
\item We propose a dynamic power consumption model for our non-cooperative
and cooperative schemes, where the power consumption of a BS changes
with its load, i.e., the number of users associated with it.
\item We evaluate the performance of the proposed schemes in terms of coverage
probability, mean achievable rate and energy efficiency by using tools
from stochastic geometry. Expressions of each metric are obtained.
And they are analyzed under different system parameters by varying
the $\textrm{SINR}$ thresholds, BS densities and transmit power. 
\item We show that the proposed SBS cooperation scheme is able to offload
more users from the MBS tier. Meanwhile, it can offer better coverage
and achievable rate for a typical user in a more energy efficient
way. Benefits of deploying more SBSs can be seen from our discussion.
\end{itemize}
The rest of this paper is organized as follows. A tractable model
for a downlink 2-tier network is presented in Section \ref{sec:model}.
And then we propose the offloading scheme and power consumption model.
In Section \ref{sec:-1}, we derive expressions of the coverage probabilities,
as well as distance analysis in the non-cooperative and cooperative
model. Then on the basis of coverage probability, average rate and
energy efficiency are obtained. Simulations are conducted in Section
\ref{sec:Numerical-Results} to show the performance comparison between
our cooperative model and non-cooperative model. Section \ref{sec:Conclusions}
concludes the paper and points out the future prospects.

\section{System Model\label{sec:model}}

\begin{figure}[tbh]
\begin{centering}
\includegraphics[scale=0.46]{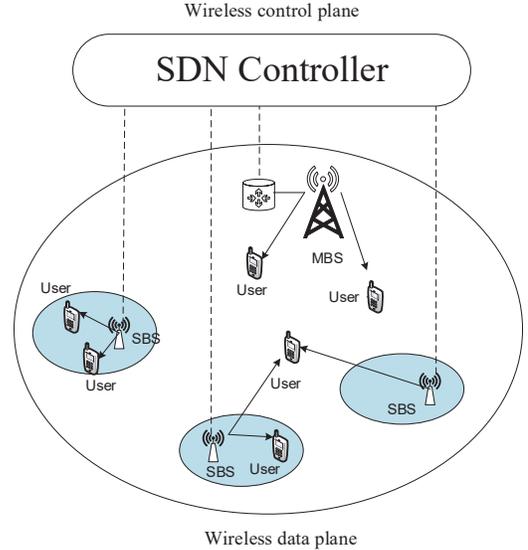}
\par\end{centering}
\centering{}\caption{\label{fig:Heterogenous Network}A software defined heterogeneous
network with small cell cooperation}
\end{figure}

We consider a software defined HetNet composed of two independent
tiers of network, i.e., the macro BS network and the small cell BS
network. Both tiers are independent with different deployment densities
and transmit powers. The BSs belonging to the MBS and SBS tier have
transmit powers $P_{m}$, $P_{s}$ and follow homogeneous Poisson
point processes (PPPs) $\Phi_{m},\Phi_{s}\in\mathbb{R^{\mathrm{2}}}$
with densities $\lambda_{m}$, $\lambda_{s}$ respectively. All the
BSs and users are assumed to be equipped with a single antenna. Without
any loss of generality, we focus on a typical user at the origin.
A simple case of a software defined 2-tier HetNet composed of a single
MBS and multiple SBSs is illustrated in Fig. \ref{fig:Heterogenous Network},
where users can be served by BSs from the two tiers. All BSs are connected
to the SDN controller by wireless links. Thus the control plane and
data plane are separated. All connections are configured by the OpenFlow
protocol, which is proposed to standardize the communications between
the data plane and control plane \cite{chen2015software}. Through
these links, BSs can transmit the related state information to the
SDN controller which sends the control information back to BSs. In
this case, applications, such as cooperation and radio resource allocation
are managed by the SDN controller in the control plane. It is similar
to the measurement flows and control flows presented in \cite{li2015unified},
that are used to collect information and control underlying hardware
and software respectively. Here, it is feasible for a user to simultaneously
connect to several cells in HetNets. We consider an orthogonal frequency
division multiple access (OFDMA) system adopted for BSs. It means
that no intra-cell interference exists, but users will suffer interferences
from other BSs in both tiers.

Let $x_{i,j}$ be the location of the $j$-th BS in tier $i$ and
$r_{i,j}$ be the distance from $x_{i,j}$ to the typical user, $h_{i,j}$
be the corresponding channel coefficient. Here, $i\in\left\{ s,m\right\} $
and $s$, $m$ denote the SBS tier and MBS tier respectively. In this
paper, we consider a Rayleigh fading model to characterize the channel
fading, i.e., $h_{i,j}\sim\exp(1)$. $\alpha>2$ is the path loss
exponent for both tiers.

\subsection{Small Cell Offloading\label{subsec:Small-Cell-Offloading}}

In the HetNet, the deployment of small cell BSs is able to offload
users from the MBSs. It can ease the traffic burden of the MBSs and
improve the system's performance to some extent. For example, CRE
of the small cells is a common way to offload users from the MBS tier
and increases the coverage. In any case, the existing study focuses
on the offloading ability of a single small cell. Besides, MBS can
reduce its dynamic power consumption when it is lightly loaded. Offloading
users to SBSs when they can offer better communication quality can
reduce the load of MBS and possibly reduce the energy consumption
of the entire network. In our study, we analyze the offloading performance
of the cooperation of small cells. For simplicity, the resource partition
between different tiers is not considered. 

In this scenario, a user is allowed to access any tier's BSs because
of open access. We consider a cell association based on maximum received-signal-strength
(RSS). Covered by the closest MSB and SBS, the user will choose one
that offers higher RSS as its serving BS \cite{elsawy2013hetnets}.
Thus a user is always associated to either an MBS or an SBS.

\subsubsection{Non-cooperative model}

A user will associate with the BS that results in the highest RSS.
As the BSs belonging to the same tier have the same transmit power,
it means a user will choose its closest MBS or SBS as its serving
BS. It's called the non-cooperative operation. We use association
probability to measure the traffic offloading.
\begin{lem}
\label{lem1:assoc pro} The probability that a user associates with
SBS tier can be expressed as

\begin{equation}
\mathcal{P_{\textnormal{sbs\_no}}}=\frac{1}{1+\frac{\lambda_{m}}{\lambda_{s}}\left(\frac{P_{m}}{P_{s}}\right)^{\frac{2}{\alpha}}}.\label{eq1:asPno}
\end{equation}
\end{lem}
\begin{IEEEproof}
$P_{m}r_{m}^{-\alpha}$ and $P_{s}r_{s}^{-\alpha}$ are the RSS received
by the typical user from its nearest MBS and SBS respectively. The
user will access to the tier which can offer it stronger RSS. Therefore

\begin{eqnarray*}
\mathcal{\mathcal{P_{\textnormal{sbs\_no}}}} & = & 1-\mathbb{P}\left(\mathnormal{P_{m}r_{m}^{-\alpha}>P_{s}r_{s}^{-\alpha}}\right)\\
 & = & 1-\mathbb{E_{\mathnormal{r_{m}}}\left[\mathbb{P}\left(\mathnormal{r_{s}}>\left(\mathnormal{r_{m}^{\alpha}P_{s}/P_{m}}\right)^{\mathrm{1/\alpha}}\right)\right]}\\
 & = & 1-\int_{0}^{\infty}\mathbb{P}\left(\mathnormal{r_{s}}>\left(\mathnormal{r^{\alpha}P_{s}/P_{m}}\right)^{\mathrm{1/\alpha}}\right)\mathnormal{f_{r_{m}}\left(r\right)\textrm{d}r},
\end{eqnarray*}
where $r_{m}$ and $r_{s}$ are the distances from the typical user
to its nearest MBS and SBS and its probability density functions (PDFs)
are $f_{r_{m}}\left(r\right)$ and $f_{r_{s}}\left(r\right)$. It
is known that the null probability of the 2-D homogeneous PPP with
density $\lambda$ in an area $A$ is $\exp\left(-\lambda A\right)$,
so $\mathnormal{r_{s}}>\left(\mathnormal{r_{m}^{\alpha}P_{s}/P_{m}}\right)^{\mathrm{1/\alpha}}$
means that there's no BS in the circle whose radius is $\left(\mathnormal{r_{m}^{\alpha}P_{s}/P_{m}}\right)^{\mathrm{1/\alpha}}$
in the small cell tier. Thus, 
\begin{equation}
\mathbb{P}\left(\mathnormal{r_{s}}>\left(\mathnormal{r^{\alpha}P_{s}/P_{m}}\right)^{\mathrm{1/\alpha}}\right)=\textrm{e}^{\left(-\lambda_{s}\pi r^{2}\left(P_{s}/P_{m}\right)^{2/\alpha}\right)},\label{eq3:}
\end{equation}
and it is proved in \cite{baumstark2007some} that 
\begin{equation}
f_{r_{m}}\left(r\right)=\frac{\textrm{d}\left(1-\mathbb{P}\left[\mathnormal{r_{m}>r}\right]\right)}{\textrm{d}r}=2\pi\lambda_{m}r\textrm{e}^{(-\pi\lambda_{m}r^{2})}.\label{eq:fr}
\end{equation}

Then by solving the exponential integral, it can be simplified as
\eqref{eq1:asPno}. 
\end{IEEEproof}
From the above, we can see that in order to change the load of each
tier, it just needs to adjust the BS density and power ratio $\lambda_{m}/\lambda_{s}$
and $P_{m}/P_{s}$. Although deploying more SBSs will bring extra
costs, the tendency of small cell densification has proved that the
capacity of heterogeneous networks can be dramatically improved. Comparing
with the method of adjusting transmit powers, deploying more SBSs
seems to be a more reasonable way to offload users from MBS. 

\subsubsection{Cooperative model}

To meet the exponential growth of traffic demands, small cell densification
is a promising way. As small cells are getting closer, cooperation
can benefit more. In our proposed network model, the SDN controller
coordinates the cooperation among SBSs by gathering information, such
as channel states, BSs and users. It decouples processing from transmission
within the data plane to make cooperation practically feasible. We
call the cooperative small cells an SBS cluster. In the cooperative
operation, the cooperative cluster consists of $k$ closest small
cells, denoted by $\mathcal{C\subset\mathsf{\Phi_{\mathnormal{s}}}}$.
The cooperative small cells jointly transmit a message to the typical
user. In other words, the user will associate with the cooperative
cluster or the MBS depending on its RSS. The association probability
of the cooperative model is

\begin{eqnarray}
\mathcal{P_{\textrm{sbs\_co}}} & = & \mathbb{P\left(\mathnormal{\overset{{\scriptstyle x_{s,k}}}{\underset{{\scriptstyle x_{s,j}\in\mathcal{C}}}{\sum}}P_{s}r_{s,j}^{-\alpha}>P_{m}r_{m}^{-\alpha}}\right)}\nonumber \\
 & = & \int\limits _{\stackrel{{\scriptstyle 0<r_{s,1}<r_{s,2...}}}{r_{s,k-1}<r_{s,k}<\infty}}\textrm{e}^{\left(-\lambda_{m}\pi\eta^{2/\alpha}\right)}f_{\Gamma}\left(\textbf{r}\right)\textrm{d}\textbf{r\textnormal{,}}\label{eq15:asPco}
\end{eqnarray}
where $\eta=P_{m}/\left(P_{s}\overset{{\scriptstyle x_{s,k}}}{\underset{{\scriptstyle x_{s,j}\in\mathcal{C}}}{\sum}}r_{s,j}^{-\alpha}\right)$
and $r_{s,j}$ denotes the distance of the $j$-th closest small cell
to the typical user.

It can be proved by referring to Lemma \ref{lem1:assoc pro}. $\mathbf{r}=[r_{s,1},r_{s,2},....,r_{s,k-1},r_{s,k}]$
is the set of the distances of the $k$ closest small cells to the
typical user. 
\begin{lem}
\label{lem2:joint fr} The joint PDF of $r_{s,j}$ is
\begin{equation}
f_{\Gamma}\left(\textbf{r}\right)=\left(2\pi\lambda_{s}\right)^{k}\textrm{e}^{\left(-\lambda_{s}\pi r_{s,k}^{2}\right)}\overset{{\scriptstyle x_{s,k}}}{\underset{{\scriptstyle x_{s,j}\in\mathcal{C}}}{\prod}}r_{s,j}.\label{eq:joint fr}
\end{equation}
\end{lem}
\begin{IEEEproof}
We start from the first and second closest neighbor SBSs. We define
$f_{r_{s,2}\mid r_{s,1}}\left(r_{s,2}\mid r_{s,1}\right)$ as the
conditional PDF of the distance of the second closest SBS. Around
the typical user, with radiuses of $r_{s,1}$ and $r_{s,2}$, it forms
an annulus. In analogy to \eqref{eq:fr}, the probability that there's
no SBS in this annulus is $\exp\left(-\lambda_{s}\pi\left(r_{s,2}^{2}-r_{s,1}^{2}\right)\right)$.
So,

\[
f_{r_{s,2}\mid r_{s,1}}\left(r_{s,2}\mid r_{s,1}\right)=2\pi\lambda_{s}\textrm{e}^{\left(-\lambda_{s}\pi\left(r_{s,2}^{2}-r_{s,1}^{2}\right)\right)}r_{s,2},
\]
and by using Bayes formula, we have

\begin{eqnarray}
f_{r_{s,2},r_{s,1}}\left(r_{s,2},r_{s,1}\right) & = & f_{r_{s,2}\mid r_{s,1}}\left(r_{s,2}\mid r_{s,1}\right)f_{r_{s}}\left(r\right)\nonumber \\
 & = & \left(2\pi\lambda_{s}\right)^{2}\textrm{e}^{\left(-\lambda_{s}\pi r_{s,2}^{2}\right)}r_{s,1}r_{s,2}.\nonumber \\
\end{eqnarray}

In a similar way, we can easily obtain the joint PDF of the $k$-th
closest cooperative SBSs expressed as \eqref{eq:joint fr}.
\end{IEEEproof}
It's obvious that the SBS cluster can offer stronger RSS comparing
with the single SBS association. So it will offload more users to
the small cell tier. But it adds extra complexity to both the transmit
side and receive side.

So in the non-cooperative model, we define $\mathcal{\mathrm{\mathnormal{A}}_{\mathrm{m}}}$
($\mathcal{\mathrm{\mathnormal{A}}_{\mathrm{s}}}$) as the event that
the user connects to an MBS (SBS). In the cooperative model, we define
$\mathcal{\mathrm{\mathnormal{B}}}_{m}$ ($\mathcal{\mathrm{\mathnormal{B}}}_{s}$)
as the event that the user connects to an MBS (SBS cluster). 

Small cell cooperation will offload more users from the MBS tier and
provide better coverage performance compared with a single small cell.
The communication quality will improve dramatically especially for
the cell-edge users. In addition, cooperation is also an effective
means to reduce interferences. However, cooperation is a complex work
for small cells to perform, and deploying more cells causes additional
costs in the future. 

\subsection{Power Consumption Model\label{subsec:Power-Consumption-Model}}

We consider the power consumption in a Voronoi cell \cite{baccelli2015stochastic}
consisting of an MBS and several users. A Voronoi cell associated
with a given MBS is the set of all points in $\mathbb{R^{\mathrm{2}}}$
which are closer to it than to any other MBSs. The power consumption
is load-dependent for MBSs, and to a lesser extent for SBSs \cite{auer2011much}.
We introduce SDN controller to manage the power control of BSs. Utilizing
the load information of the BS, the SDN controller notifies the BS
to turn on/off some channels for dynamically adjusting the BS power
consumption.

\subsubsection{Non-cooperative model}

The power consumption of an MBS is proportional to its current load,
which is the number of users associated with it. The total power consumption
of the MBS under non-cooperative model is 
\begin{equation}
P_{\textrm{mbs\_no}}=P_{ms}+nP_{max}\left(\mathcal{\mathrm{1-}P}_{\textnormal{sbs\_no}}\right)/N,\label{eq:pm-no}
\end{equation}
where $P_{ms}$ denotes the static power expenditure, such as processing
unit and radio module. $P_{max}$ is the maximum output power. $n$
is the number of users in the Voronoi cell, $N$ is the maximum number
of the users when the MBS is full loaded and quality of communication
can also be guaranteed in the meantime. $n\left(\mathcal{\mathrm{1-}P}_{\textnormal{sbs\_no}}\right)$
denotes the number of users connecting to the MBS. The output power
is determined by the ratio $n\left(\mathcal{\mathrm{1-}P}_{\textnormal{sbs\_no}}\right)/N$. 

Since the power consumption of small cell is low and most part of
it is static consumption which is $P_{s}$, we assume the energy consumption
of the small cells which the users in the Voronoi cell associate with
is
\begin{equation}
P_{\textrm{sbs\_no}}=n\mathcal{P}_{\textnormal{sbs\_no}}P_{s}.\label{eq:ps-no}
\end{equation}

We define the total power consumption of the system as the energy
needed to serve all the users in the Voronoi cell, which is $P_{\textrm{no}}=P_{\textrm{mbs\_no}}+P_{\textrm{sbs\_no}}$.

\subsubsection{Cooperative model}

In the cooperative model, as more users choose the small cell tier,
the number of users connecting to MBS turns to $n\left(\mathrm{1-}P_{\textrm{sbs\_co}}\right)$,
so the total energy consumption of the MBS is
\begin{equation}
P_{\textrm{mbs\_co}}=P_{ms}+nP_{max}\mathcal{\left(\mathrm{1-}P_{\textrm{sbs\_co}}\right)}/N.\label{eq:pm-co}
\end{equation}

If $k$ small cells cooperate, that means a user is served by $k$
BSs simultaneously. Due to the low power consumption of small cell,
it can be assumed that there are totally $knP_{\textrm{sbs\_co}}$
small cells serving for users in a Voronoi cell. So the energy consumption
of the small cells is
\begin{equation}
P_{\textrm{sbs\_co}}=kn\mathcal{P}_{\textrm{sbs\_co}}\left(P_{s}+P_{bkh}\right).\label{eq:ps-co}
\end{equation}

$P_{bkh}$ is the backhaul power consumption for each cooperative
SBS. To serve a common user, the cooperative SBSs have to share data
with each other, so the backhaul overhead has to be considered. When
traffic load of MBS is offloaded to SBS, the dynamic power consumption
of MBS decreases. This will be more remarkable in the cooperative
model. 

So the total power consumption for the cooperative model is $P_{\textrm{co}}=P_{\textrm{mbs\_co}}+P_{\textrm{sbs\_co}}$.

\section{SINR Coverage and Energy Efficiency \label{sec:-1}}

In this section, we analyze the offloading performance through three
main metrics: SINR coverage, user rates, and energy efficiency. First,
we will discuss the SINR for downlink transmission at a typical user
under different association strategies. Then, we derive the analytical
expressions for the coverage probability and the average ergodic rate.
On the basis of the above analysis, we can get the energy efficiency
of the system under both cooperative and non-cooperative schemes.

\subsection{SINR Coverage Probability}

The received signal at a typical user can be written as 
\begin{equation}
\underset{\mathit{{\scriptstyle x_{i,j}\in\mathcal{B}}}}{\sum}\frac{\left(Ph_{i,j}\right)^{1/2}}{r_{i,j}^{\alpha/2}}X+\underset{x_{i,j}\notin\mathit{\mathbb{\mathcal{B}}}}{\sum}\frac{\left(Ph_{i,j}\right)^{1/2}}{r_{i,j}^{\alpha/2}}Y+Z,
\end{equation}
where $P$ is the transmit power of the serving BS. $\mathcal{B}$
denotes the set of BSs that the user associate with, it may be an
MBS, an SBS or an SBS cluster. $X$ denotes the input symbol that
is sent by the associated BSs. So the first sum denotes the useful
signal from the associated BSs. $Y$ denotes the input symbol sent
by the BSs that do not belong to $\mathcal{B}$. So the second sum
denotes the interference, including the inter-tier and intra-tier
interference. $Z$ is a circular-symmetric zero-mean complex Gaussian
random variable with variance $\sigma^{2}$, which models the additive
white noise. 

Thus, from \cite{nigam2014coordinated} and \cite{sakr2014location},
the received instantaneous SINR at a typical user is given by
\begin{equation}
\mathrm{SINR\left(\mathcal{B}\right)=\frac{\mid\underset{\mathit{{\scriptstyle x_{i,j}\in\mathcal{B}}}}{\sum}\mathnormal{\left(Ph_{i,j}\right)^{1/2}r_{i,j}^{-\alpha/\mathrm{2}}}\mid^{2}}{\mathnormal{\mid\underset{\mathit{{\scriptstyle x_{i,j}\notin\mathcal{B}}}}{\sum}\left(Ph_{i,j}\right)^{1/2}r_{i,j}^{-\alpha/2}\mid^{2}+\sigma^{2}}}}.
\end{equation}

The coverage probability is defined as the probability that the received
SINR is greater than a threshold. For a given SINR threshold $\theta$,
the coverage probability $\mathcal{P}_{n}$ at the typical user can
be expressed as
\begin{equation}
\mathcal{P}_{n}=\mathbb{P\left(\mathrm{SINR>\theta}\right)},\label{eq:p-cov}
\end{equation}
and $\mathcal{P}_{n}$ is relevant to the user's association policy. 
\begin{lem}
\label{lem3:con fr}To connect the link between the coverage and the
association policy, we analyze the PDFs of the distance between a
typical user and its serving BS or BSs, i.e., $f_{R}\left(r\right)$.
In the non-cooperative model, when the user connects to an MBS or
an SBS, the PDFs are respectively

\begin{equation}
f_{Rm}\left(r\right)=\frac{2\pi\lambda_{m}r}{1-\mathcal{P_{\textnormal{sbs\_no}}}}\textrm{e}^{\left(-\pi r^{2}\left(\lambda_{m}+\lambda_{s}\left(P_{s}/P_{m}\right)^{2/\alpha}\right)\right)},\label{eq:fr1-con}
\end{equation}

and

\begin{equation}
f_{Rs}\left(r\right)=\frac{2\pi\lambda_{s}r}{\mathcal{P_{\textnormal{sbs\_no}}}}\textrm{\textrm{\textrm{e}}}^{\left(-\pi r^{2}\left(\lambda_{s}+\lambda_{m}\left(P_{m}/P_{s}\right)^{2/\alpha}\right)\right)}.\label{eq:fr2-con}
\end{equation}

As the association probability changes, the distance distribution
changes as well. In the cooperative model, the PDF of the distance
between the user and the MBS is 
\begin{equation}
f_{Rcm}\left(r\right)=\frac{f_{r_{m}}\left(r\right)g(r)}{1-P_{\textnormal{sbs\_co}}},\label{eq:joint f1}
\end{equation}
where $\mathbb{\mathnormal{g(r)=}P}\left(\mathnormal{\mathnormal{P_{m}r^{-\alpha}/P_{s}>\mathnormal{\overset{{\scriptstyle x_{s,k}}}{\underset{{\scriptstyle x_{s,j}\in\mathcal{C}}}{\sum}}r_{s,j}^{-\alpha}}}}\right)$
is the function of $r$. $f_{r_{m}}\left(r\right)$ is given as \eqref{eq:fr}.

If the user connects to an SBS cluster, the joint PDF is
\begin{eqnarray}
f_{Rcs}\left(\textbf{r}\right) & = & \frac{1}{P_{\textnormal{sbs\_co}}}\textrm{e}^{\left(-\pi\lambda_{m}\eta^{2/\alpha}\right)}f_{\Gamma}\left(\textbf{r}\right).\label{eq:frc2}
\end{eqnarray}
\end{lem}
\begin{IEEEproof}
See Appendix \ref{sec:app a}. 
\end{IEEEproof}
\begin{rem}
\label{rem:pmco}Note that there is no general analytical expression
for $g(r)$ in \eqref{eq:joint f1}, we can derive an exact result
for a special case for simplicity, as shown later in our simulation.
Here, we consider two SBSs cooperation and $\alpha=4$. So we have
\begin{equation}
g(r)=\int_{0}^{\frac{\pi}{4}}\frac{\pi\lambda_{s}\sin^{-1}\varphi+\omega r^{-2}}{\left(\cos\varphi r^{-1}\right)^{2}\omega}\textrm{e}^{\left(-\frac{\pi\lambda_{s}r^{2}}{\omega\sin\varphi}\right)}\textrm{d}\varphi,\label{eq:gr}
\end{equation}
where $\omega=\sqrt{P_{m}/P_{s}}$.
\end{rem}
\begin{IEEEproof}
See Appendix \ref{sec:PROOF-OF-REMARK}.
\end{IEEEproof}
\begin{thm}
\label{thm:covp}The coverage probabilities for a typical user associated
with an MBS, an SBS in non-cooperative model are

\begin{equation}
\mathcal{P}_{n}\left(A_{m}\right)=\int_{r>0}\left[\textrm{e}^{\left(-P_{m}^{-1}\theta r^{\alpha}\sigma^{2}\right)}\mathcal{L_{\mathnormal{I}}\left(\mathnormal{P_{m}^{-1}}\mathnormal{\theta r^{\alpha}}\right)}\right]f_{Rm}\left(r\right)\textrm{d}r,\label{eq:pn1}
\end{equation}

\begin{equation}
\mathcal{P}_{n}\left(A_{s}\right)=\int_{r>0}\left[\textrm{e}^{\left(-P_{s}^{-1}\theta r^{\alpha}\sigma^{2}\right)}\mathcal{L_{\mathnormal{I}}\left(\mathnormal{\mathnormal{P_{s}^{-1}}\theta r^{\alpha}}\right)}\right]f_{Rs}\left(r\right)\textrm{d}r,\label{eq:pn2}
\end{equation}
where distance distributions are given in Lemma \ref{lem3:con fr}.
And

\begin{equation}
\mathcal{L}_{\mathnormal{I}}\left(\mathnormal{s}\right)=\overset{}{\underset{i\in\left\{ s,m\right\} }{\prod}}\textrm{e}^{\left(-2\pi\lambda_{i}\left(sP_{i}\right)^{2/\alpha}\mathcal{F\left(\mathnormal{\left(sP_{i}\right)^{\mathnormal{-1/\alpha}}d_{i},\alpha}\right)}\right)}.
\end{equation}
It denotes the Laplace transform of interference $I$. $d_{i}$ is
the minimum distance between the user and its nearest interfering
BS in tier $i$, $i\in\left\{ s,m\right\} .$ It means all interfering
BSs are distributed outside the circle with the radius $d_{i}$. For
\eqref{eq:pn1}, we have $d_{m}=r$ and \textup{$d_{s}=\omega^{-2/\alpha}r$.}
For \eqref{eq:pn2}, we have $d_{m}=\omega^{2/\alpha}r$ and \textup{$d_{s}=r$}.
And

\begin{equation}
\mathcal{F\left(\mathnormal{y,\alpha}\right)=\intop}_{y}^{\infty}\frac{\mu}{1+\mu^{\alpha}}\textrm{d}\mu.\label{eq20:f(y,a)}
\end{equation}
\end{thm}
\begin{IEEEproof}
See Appendix \ref{sec:app b}.
\end{IEEEproof}
Since the association methods are mutually exclusive, by using the
law of total probability, we obtain the overall coverage probability
of the non-cooperative model as follow:

\begin{equation}
\mathcal{P}_{\textrm{no}}=\left(1-\mathcal{P}_{\mathnormal{\textrm{sbs\_no}}}\right)\mathcal{P}_{n}\left(A_{m}\right)+\mathcal{P}_{\mathnormal{\textrm{sbs\_no}}}P_{n}\left(A_{s}\right),\label{eq23:overall no}
\end{equation}
where $\mathcal{P}_{\mathnormal{\textrm{sbs\_no}}}$ is given in \eqref{eq1:asPno}.

In view of the big advantage in transmit power, MBS can offer better
coverage performance than the SBS under the same threshold $\theta$.
In addition, an SBS user may be subjected to more cross-tier interference. 
\begin{rem}
We can get the closed-form expressions for specific values of the
integral function \eqref{eq20:f(y,a)} \cite{nigam2014coordinated}.
Then we can get $\mathcal{F}\mathnormal{\left(y,4\right)}=\frac{1}{2}\tan^{-1}\left(y^{-2}\right).$
Note that in the interference-limited network, when $\sigma^{2}$
is small enough, the first exponential term in \eqref{eq:pn1} and
\eqref{eq:pn2} approaches $1$. Here, in consideration of the situation
and conciseness of the numerical analysis, assuming $\alpha=4$ and
$\sigma^{2}=0$, we have the result $\mathcal{P}_{n}\left(A_{m}\right)=\mathcal{P}_{n}\left(A_{s}\right)$.
And the total outage probability \eqref{eq23:overall no} simplifies
to 

\begin{equation}
\mathcal{P}_{\textrm{no}}=\frac{1}{1+\sqrt{\theta}\tan^{-1}\left(\sqrt{\theta}\right).}\label{eq:pno}
\end{equation}
\end{rem}
For this special case, the total outage probability in the non-cooperative
model is just the function of threshold $\theta$ and there exists
a negative correlation between them. This is the same as the result
in \cite{dhillon2012modeling}, where the coverage probability has
nothing to do with the number of the tiers or their densities and
transmit powers in an interference-limited network. The adjustment
in the properties of the BSs is helpless to improve the coverage performance.
In other words, we can deploy more BSs to improve the throughput and
ignoring the interference they cause.
\begin{thm}
\label{thm:cooperative covp}In the cooperative model, when the user
associates with an MBS we can obtain its coverage probability $\mathcal{P}_{n}\left(B_{m}\right)$
by substituting $f_{Rcm}\left(r\right)$ for $f_{Rm}\left(r\right)$
in \eqref{eq:pn1}. When the user associates with an SBS cluster,
its coverage probability $\mathcal{P}_{\mathnormal{n}}\left(B_{s}\right)$
is shown as (25). Same as the coverage probabilities in non-cooperative
model, for $\mathcal{P}_{n}\left(B_{m}\right)$ we have $d_{m}=r$
and \textup{$d_{s}\thickapprox\omega^{-2/\alpha}r$}. And for $\mathcal{P}_{n}\left(B_{s}\right)$,
we have $d_{m}=\eta^{1/\alpha}$ and $d_{s}=r_{s,k}$.
\end{thm}
\begin{IEEEproof}
See Appendix \ref{sec:app b}.
\end{IEEEproof}
\begin{figure*}[tbh]
\begin{equation}
\mathcal{P}_{n}\left(B_{s}\right)=\int\limits _{\stackrel{{\scriptstyle 0<r_{s,1}<r_{s,2...}}}{r_{s,k-1}<r_{s,k}<\infty}}\left[\exp\left(\frac{-\theta\sigma^{2}}{\overset{{\scriptstyle x_{s,k}}}{\underset{{\scriptstyle x_{s,j}\in\mathcal{C}}}{\sum}}P_{s}r_{s,j}^{-\alpha}}\right)\mathcal{L_{\mathnormal{I}}\left(\frac{\mathnormal{\theta}}{\mathnormal{\overset{{\scriptstyle x_{s,k}}}{\underset{{\scriptstyle x_{s,j}\in\mathcal{C}}}{\sum}}P_{s}r_{s,j}^{-\alpha}}}\right)}\right]f_{Rcs}\left(\textbf{r}\right)\textrm{d}\textbf{r}\label{eq:pcovco}
\end{equation}

\rule[0.5ex]{1\textwidth}{0.5pt}
\end{figure*}
The overall coverage probability of the user in the cooperative model
is

\begin{equation}
\mathcal{P}_{\textrm{co}}=\left(1-\mathcal{P}_{\mathnormal{\textrm{sbs\_co}}}\right)\mathcal{P}_{n}\left(B_{m}\right)+\mathcal{P}_{\mathnormal{\textrm{sbs\_co}}}\mathcal{P}_{n}\left(B_{s}\right).\label{eq24:overall co}
\end{equation}

When $k=1$, \eqref{eq:pcovco} degrades into \eqref{eq:pn2}. In
the SBS tier, if the user associates with only one BS, its second
nearest neighbor BS becomes its strongest interference. Same for the
$k$-th nearest neighbor. So if we bring in cooperation, the strongest
interference turns into useful signal. As $k$ increases, this advantage
will be more obvious. Even two SBSs cooperate, it may have a better
coverage performance than the MBS user. For the cell-edge users who
are far from the MBS, a single SBS is not able to satisfy its requirements
for communication. Cooperation could solve this problem. Because cooperation
needs BSs to exchange data with each other, simply increasing $k$
is not an optimal solution. And it is also not practical for the mobile
users to connect to too many BSs at the same time.

\subsection{Mean Achievable Rate}

With the conditional coverage probability above, we can easily obtain
the mean achievable rates (measured in (bit/sec/Hz)) of non-cooperative
and cooperative models respectively. 
\begin{thm}
The achievable rate of a typical user can be expressed as
\begin{equation}
\tau=\frac{1}{\ln2}\int_{0}^{\infty}\mathcal{P}_{n}\frac{1}{1+\theta}\mathbb{\mathnormal{\textrm{d}\theta}},\label{eq:rate}
\end{equation}
where $\mathcal{P}_{n}$ is the coverage probability of the given
user. After substituting $\mathcal{P}_{n}\left(A_{m}\right),\mathcal{P}_{n}\left(A_{s}\right),\mathcal{P}_{n}\left(B_{m}\right)$
and $\mathcal{P}_{n}\left(B_{s}\right)$ into \eqref{eq:rate}, we
can get the achievable rates when user associates to the corresponding
BS, which are $\tau_{\textrm{mbs}}$, $\tau_{\textrm{sbs}}$ and $\tau_{\textrm{mbs\_co}}$,
$\tau_{\textrm{mbs\_co}}$.
\end{thm}
\begin{IEEEproof}
The mean achievable rate of the typical user is defined as 
\begin{eqnarray*}
\tau & = & \mathbb{E}\left[\mathnormal{\log\mathnormal{_{2}}\left(\mathnormal{\textrm{1}}+\mathrm{SINR}\right)}\right]\\
 & = & \frac{1}{\ln2}\mathbb{E\left[\mathnormal{\ln\left(\mathnormal{\textrm{1}}+\mathrm{SINR}\right)}\right]}\\
 & = & \frac{1}{\ln2}\int_{0}^{\infty}\mathbb{P}\left(\mathrm{SINR}>\textrm{e}^{t}-1\right)\textrm{d}t\\
 & = & \frac{1}{\ln2}\int_{0}^{\infty}\mathbb{P}\left(\mathrm{SINR}>\theta\right)\frac{1}{1+\theta}\textrm{d}\theta.
\end{eqnarray*}

Here using the change of variables and the definition of the coverage
probability \eqref{eq:p-cov} we can obtain the final result.
\end{IEEEproof}
According to the mutual independence of different association strategy,
by using the law of total probability, we get the mean achievable
rate for a user in non-cooperative model as 
\begin{equation}
\tau_{\textrm{no}}=\left(1-\mathcal{P}_{\mathnormal{\textrm{sbs\_no}}}\right)\tau_{\mathnormal{\textrm{mbs}}}+\mathcal{P}_{\mathnormal{\textrm{sbs\_no}}}\tau_{\mathnormal{\textrm{sbs}}}.
\end{equation}

Similarly, the mean achievable rate in the cooperative model is

\begin{equation}
\tau_{\textrm{co}}=\left(1-\mathcal{P}_{\mathnormal{\textrm{sbs\_co}}}\right)\tau_{\mathnormal{\textrm{mbs\_co}}}+\mathcal{P}_{\mathnormal{\textrm{sbs\_co}}}\tau_{\mathnormal{\textrm{sbs\_co}}}.
\end{equation}

\subsection{Energy Efficiency}

Energy efficiency is one of the key performance indicators for the
proposed model. We define it as the ratio of throughput to the energy
consumption of the network (usually measured in (bits/J)). It is assumed
that BSs in different tiers share the same frequency bandwidth $B$.
Since we have derived the achievable rate $\tau$, the throughput
of a typical user is $\tau B$. For a Voronoi cell, it includes all
the BSs that serve the users in the cell. Numbers of the users and
BSs are given in \ref{subsec:Power-Consumption-Model}. The total
system throughput of the non-cooperative model and cooperative model
are respectively derived as

\begin{equation}
R_{\textrm{no}}=n\tau_{\textrm{no}}B,
\end{equation}
and
\begin{equation}
R_{\textrm{co}}=n\tau_{\textrm{co}}B.
\end{equation}
Thus, the expressions of the corresponding energy efficiency of the
non-cooperative model and cooperative model are 

\begin{equation}
\mathcal{E}_{\mathrm{\textrm{no}}}=R_{\textrm{no}}/P_{\textrm{no}},
\end{equation}
and

\begin{equation}
\mathcal{E}_{\textrm{\ensuremath{\mathrm{\textrm{co}}}}}=R_{\textrm{co}}/P_{\textrm{co}}
\end{equation}
 respectively.

\section{Numerical Results\label{sec:Numerical-Results}}

In this section, the traffic offloading performance with cooperative
and non-cooperative models has been compared by numerical results.
Without loss of generality, two small cells cooperative scenarios
are analyzed and $\sigma^{2}=0$ is configured in numerical simulations.
The comparison is performed in terms of coverage probability, association
probability, average achievable rate and energy efficiency. 

\subsection{Coverage Probability}

\begin{figure}[tbh]
\begin{centering}
\includegraphics[width=8cm]{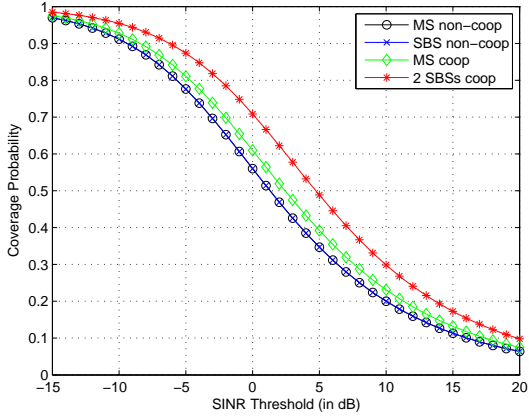}
\par\end{centering}
\caption{\label{fig:cov-thres}Coverage probabilities when the user connects
to different BSs. $\alpha=4,$ $\lambda_{m}=\left(500^{2}\pi\right)^{-1}$,
$\lambda_{s}=50\lambda_{m}$ and $P_{m}=50$, $P_{s}=1$.}
\end{figure}

Fig. \ref{fig:cov-thres} shows the coverage probability trends with
$\mathnormal{\textnormal{SINR}}$ threshold corresponding to event
$A_{m}$, $A_{s}$, $B_{m}$ and $B_{s}$. The coverage probability
decreases as the $\mathnormal{\textnormal{SINR}}$ threshold increases.
The curves of the SBS and MBS are the same in non-cooperative model.
In this interference-limited regime, the user just connects to a BS
that offers the highest RSS without differentiating between the tiers.
When two SBSs cooperates the coverage probability has been dramatically
improved and is higher than the coverage probability transmitted by
an MBS. 

\begin{figure}[tbh]
\begin{centering}
\includegraphics[width=8cm]{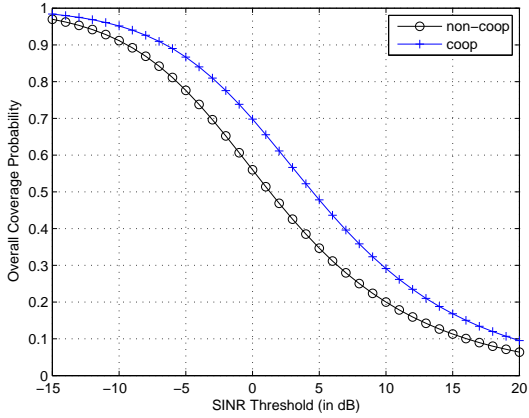}
\par\end{centering}
\caption{\label{fig:over-cov}Overall coverage probability of the non-cooperative
and cooperative model where $\lambda_{m}=\left(500^{2}\pi\right)^{-1}$,
$\lambda_{s}=50\lambda_{m}$ and $P_{m}=50$, $P_{s}=1$.}
\end{figure}

Fig. \ref{fig:over-cov} depicts the overall coverage probabilities
of the non-cooperative and cooperative model with respect to the SINR
threshold. The overall coverage probability with the non-cooperative
and cooperative model decreases with the increase of the SINR threshold.
When the SINR threshold has been fixed, the overall coverage probability
with the cooperative model is higher than that of the non-cooperative
model. It demonstrates that the cooperation can offer better coverage. 

\begin{figure}[tbh]
\begin{centering}
\includegraphics[width=8cm]{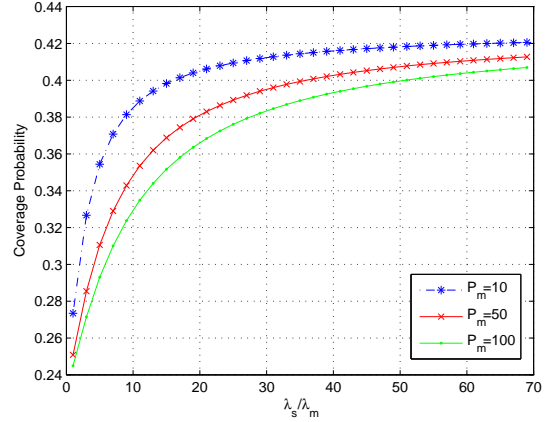}
\par\end{centering}
\caption{\label{fig:over-cov-den}Overall coverage probability of the cooperative
model under different $P_{m}$ and $\lambda_{s}$. $\lambda_{m}=\left(500^{2}\pi\right)^{-1}$
$\theta=5$ and $P_{s}=1$.}
\end{figure}

Fig. \ref{fig:over-cov-den} shows the effect of SBS density and the
transmit power of MBS on the overall coverage probability in the cooperative
model. As the SBS density grows, coverage probabilities increase and
this trend slows down gradually when $\lambda_{s}/\lambda_{m}$ is
larger than 30. That\textquoteright s because the user will associate
with the SBS cluster with a significant possibility if $\lambda_{s}$
is big enough. As the transmit power of MBS $P_{m}$ increases, there\textquoteright s
more chance that the user will associate with an MBS. But based on
the result in Fig. \ref{fig:over-cov}, the SBS cooperation offer
a better coverage. So the overall coverage probability decreases with
the increase of $P_{m}$.

\subsection{Traffic load}

\begin{figure}[tbh]
\centering{}\includegraphics[width=8cm]{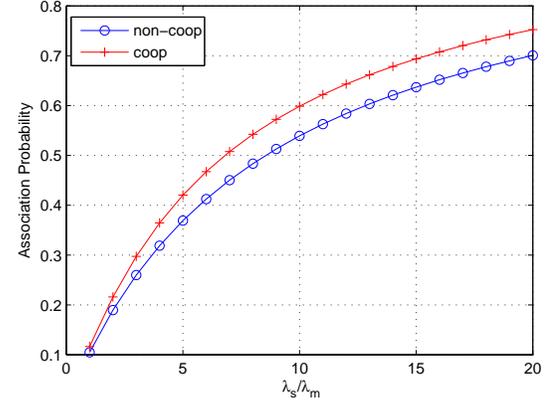}\caption{\label{fig:asP-den}Comparison of the association probability between
non-cooperative and cooperative model where $\alpha=3$, $\lambda_{m}=\left(500^{2}\pi\right)^{-1}$
and $P_{m}=50$, $P_{s}=2$.}
\end{figure}

As mentioned in \ref{subsec:Small-Cell-Offloading}, we use association
probability to evaluate small cell offloading. It can be seen from
Fig. \ref{fig:asP-den}, if small cells are deployed denser, they
will have more chance to serve the user. And the rising trends recede
as $\lambda_{m}$ increases. The association probability of cooperative
model always outperforms the non-cooperative model owe to the large
RSS of the SBS cooperation.

\subsection{Energy Efficiency}

\begin{figure}[tbh]
\begin{centering}
\includegraphics[width=8cm]{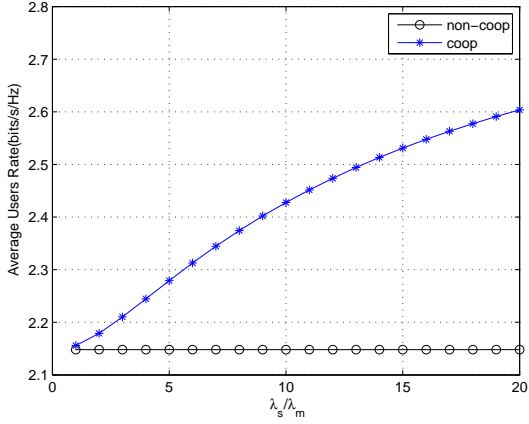}
\par\end{centering}
\caption{\label{fig:user rate}Mean achievable rate of the non-cooperative
and cooperative model where $\alpha=4$, $\lambda_{m}=\left(500^{2}\pi\right)^{-1}$,
$P_{m}=50$, $P_{s}=2$ and $B=20\textrm{MHz}$.}
\end{figure}

Fig. \ref{fig:user rate} gives the mean achievable rate of a typical
user in the two models. We still focus on the impact of the small
cell density. According to the figure, the user rate is constant as
$\lambda_{s}$ increases. We can see from \eqref{eq:rate} that user
rate is deduced from the coverage probability. \eqref{eq:pno} is
independent with the cell densities and transmit power, so there's
no doubt that the mean achievable rate is invariant for $\lambda_{s}/\lambda_{m}$
in the non-cooperative model. In the cooperative model, the rate is
a rising curve and the trend is steady. Deploying more SBSs brings
the SBSs closer to the user, thus SINR and rate of the user becomes
higher.

\begin{figure}[tbh]
\begin{centering}
\includegraphics[width=8cm]{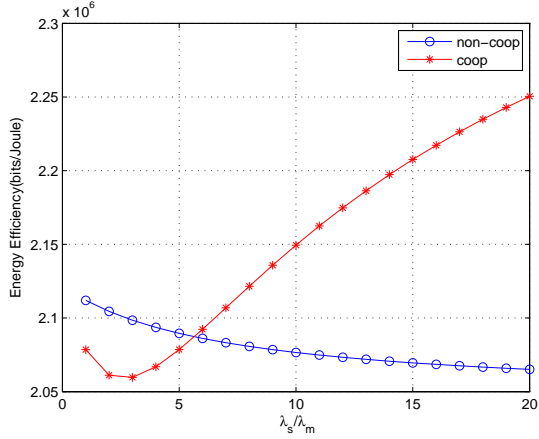}
\par\end{centering}
\caption{\label{fig:EE}The system's energy efficiency of the non-cooperative
and cooperative model where $\alpha=4$, $\lambda_{m}=\left(500^{2}\pi\right)^{-1}$,
$P_{m}=50$, $P_{s}=2$, $B=20\textrm{MHz}$ and $P_{ms}=20,$ $P_{bkh}=1$.}
\end{figure}

Fig. \ref{fig:EE} shows the energy efficiency of the non-cooperative
and cooperative model in a Voronoi cell. The shape of the curves is
as similar with that of the user rate shown in Fig. \ref{fig:user rate}.
In the non-cooperative model, the energy efficiency decreases with
the increase of the $\lambda_{s}/\lambda_{m}$. This really distracts
from our main goal. In the cooperative model, when $\lambda_{s}<3\lambda_{m}$
the energy efficiency experiences a slight decrease. When $\lambda_{s}>3\lambda_{m}$
it keeps up increase. Considering the result in Fig. \ref{fig:asP-den}
, when $\lambda_{s}$ is small, user is more likely to associate with
the MBS other than the cooperative SBSs. When $\lambda_{s}=6\lambda_{m}$,
the two curves meet and $\mathcal{E}_{\mathrm{\textrm{co}}}$ starts
to transcend $\mathcal{E}_{\textrm{\ensuremath{\mathrm{no}}}}$. Hence,
we can draw the conclusion that small cell offloading through cooperation
have better energy efficiency in a dense environment. 

In our work, we analyze $n$ users in a Voronoi cell where there\textquoteright s
one MBS and several SBSs to serve these users. The number of SBSs
is decided by $n$ and is constant, and many of the SBSs may even
locate outside the Voronoi cell. In the cooperative model, as $\lambda_{s}$
increases, there won\textquoteright t be more SBSs to serve these
$n$ users, thus the energy consumption won\textquoteright t increase
significantly. Meanwhile, an increase in $\lambda_{s}$ means the
SBSs are closer to the users, so the user rate will improve through
cooperation. These changes make EE improves as $\lambda_{s}$ increases.

\section{Conclusions\label{sec:Conclusions}}

In this paper, we have proposed a novel offloading strategy through
small cell cooperation. Using tools from stochastic geometry, a tractable
model has been proposed in the downlink HetNets. The SDN controller
has been introduced to manage the SBS cooperation and power control,
which simplifies the BS's function and burden. In proposed cooperative-model,
two or more small cells could serve a common user simultaneously.
The user connects to an MBS or an SBS cluster depending on its RSS.
In our study the power consumption of BS is determined by its load,
which is represented by the association probability. We obtain the
expressions of the overall coverage probabilities, achievable rate
for a typical user with and without cooperation. Numerical results
have shown that small cell cooperation could offload more users from
MBS tier. It can also increase the system's coverage performance.
It's been proved that when SBSs get closer, cooperation can benefit
more and thus shows the great potential of small cell densification.

\appendices{}

\section{Proof of Lemma \ref{lem3:con fr}\label{sec:app a}}

$f_{R}\left(r\right)$ is a conditional PDF under different association
circumstance. So for the MBS association in non-cooperative model,
the probability of the event of $\mathnormal{R_{m}}\mathnormal{>r}$
is 
\begin{eqnarray}
\mathbb{P}\left[\mathnormal{R_{m}}\mathnormal{>r}\right] & = & \mathbb{P}\left[\mathnormal{r_{m}}\mathnormal{>r}\mid A_{m}\right]\nonumber \\
 & = & \frac{\mathbb{P}\left[\mathnormal{r_{m}}\mathnormal{>r},A_{m}\right]}{\mathbb{P}\left[\mathnormal{A_{m}}\right]}\nonumber \\
 & = & \frac{\mathbb{P}\left[\mathnormal{r_{m}}\mathnormal{>r},\mathnormal{\mathnormal{P_{m}r_{m}^{-\alpha}>P_{s}r_{s}^{-\alpha}}}\right]}{1-\mathcal{P_{\textnormal{sbs\_no}}}}\nonumber \\
 & = & \frac{\int_{r}^{\infty}\mathbb{P}\left[\mathnormal{\mathnormal{r_{s}>r\omega^{\mathnormal{-2/\alpha}}}}\right]f_{r_{m}}\left(r\right)\textrm{d}r}{1-\mathcal{P_{\textnormal{sbs\_no}}}},\label{eq:appen1}
\end{eqnarray}
where $\mathnormal{r_{m}}$ is the distance of the nearest MBS, $\mathbb{P}\left[\mathnormal{\mathnormal{r_{s}>\left(P_{s}r^{\alpha}/P_{m}\right)^{\mathnormal{1/\alpha}}}}\right],$
$f_{r_{m}}\left(r\right)$ and $\mathcal{P_{\textnormal{sbs\_no}}}$
can be found in Lemma \ref{lem1:assoc pro}. Then we can get $f_{Rm}\left(r\right)=\frac{\textrm{d}\mathbb{P\left[\mathnormal{r_{m}}\mathnormal{>r}\mid\mathnormal{A_{m}}\right]}}{\textrm{d}r}.$
Proofs of \eqref{eq:fr2-con} is the same under the condition of $A_{s}$.

For \eqref{eq:joint f1}, 
\begin{eqnarray}
\mathbb{P}\left[\mathnormal{r_{m}}\mathnormal{>r},\mathnormal{B_{m}}\right] & = & \mathbb{P}\left[\mathnormal{r_{m}}\mathnormal{>r},\mathnormal{\mathnormal{P_{m}r_{m}^{-\alpha}>\mathnormal{\overset{{\scriptstyle x_{s,k}}}{\underset{{\scriptstyle x_{s,j}\in\mathcal{C}}}{\sum}}P_{s}r_{s,j}^{-\alpha}}}}\right]\nonumber \\
 & = & \int_{r}^{\infty}\left(\mathbb{P}\left(\mathnormal{\mathnormal{r<\eta^{\mathnormal{1/\alpha}}}}\right)\right)f_{r_{m}}\left(r\right)\textrm{d}r.
\end{eqnarray}

For \eqref{eq:frc2}, we can get joint probability
\begin{equation}
\mathbb{P}\left[\mathnormal{r_{m}}\mathnormal{>r},\mathnormal{B_{s}}\right]=\int\limits _{\stackrel{{\scriptstyle 0<r_{s,1}<r_{s,2...}}}{r_{s,k-1}<r_{s,k}<\infty}}\mathbb{P}\left[\mathnormal{r>\eta}^{1/\alpha}\right]f_{\Gamma}\left(\textbf{r}\right)\textrm{d\textbf{r}}
\end{equation}
as similar with the numerator in \eqref{eq:appen1}. After the multiple
integral we can obtain the result.

\section{Proof of Remark \ref{rem:pmco}\label{sec:PROOF-OF-REMARK}}

The proof of \eqref{eq:gr} is under the condition that $k=2$ and
$\alpha=4$. So 

\begin{eqnarray}
g\left(r\right) & = & \mathbb{P\mathnormal{\left(\omega^{\mathnormal{2}}r^{\mathnormal{-4}}>\mathnormal{r_{s,1}^{\mathnormal{-4}}}+\mathnormal{r_{s,2}^{\mathnormal{-4}}}\right)}}\nonumber \\
 & = & \mathbb{P}\mathnormal{\left(\left(\omega r^{\mathnormal{-2}}\right)^{\mathnormal{2}}>\left(\mathnormal{r_{\mathnormal{s,1}}^{\mathnormal{-2}}}\right)^{\mathnormal{2}}+\left(\mathnormal{\mathnormal{r_{s,2}^{-2}}}\right)^{\mathnormal{2}}\right)},
\end{eqnarray}
so this probability is constrained in an area which within a circle
with radius $\omega r^{-2}$ and the area $0<r_{s,1}<r_{s,2}<\infty.$
Set $a=\mathnormal{\mathnormal{r_{\mathnormal{s,1}}^{\mathnormal{-2}}}}$
and $b=\mathnormal{\mathnormal{r_{s,2}^{-2}}}$, the joint probability
of $a$ and $b$ can be expressed as
\begin{equation}
f_{a,b}\left(a,b\right)=\left|\begin{array}{c}
\textrm{J}\end{array}\right|f_{r_{s,2},r_{s,1}}\left(a^{-\frac{1}{2}},b^{-\frac{1}{2}}\right),
\end{equation}
where Jacob determinant $\mid\textrm{J}\mid=\left|\begin{array}{cc}
\frac{\textrm{d}r_{s,1}}{\textrm{d}a} & 0\\
0 & \frac{\textrm{d}r_{s,2}}{\textrm{d}b}
\end{array}\right|=\frac{1}{4}\left(ab\right)^{-\frac{3}{2}}$.

As it concerned to the circle area, we use the polar transformation,
so $a=\rho\cos\varphi$ and $b=\rho\sin\varphi$. Finally we have

\begin{equation}
g\left(r\right)=\int_{0}^{\frac{\pi}{4}}\textrm{d}\varphi\int_{0}^{\omega r^{-2}}\rho f_{a,b}\left(\rho\cos\varphi,\rho\sin\varphi\right)\textrm{d}\rho,
\end{equation}
after solving the inner integration, we can obtain \eqref{eq:gr}. 

\section{Proof of Theorem \ref{thm:covp} and Theorem \ref{thm:cooperative covp}\label{sec:app b}}

The proof of \eqref{eq:pn1}, \eqref{eq:pn2} and \eqref{eq:pcovco}
are similar. We take \eqref{eq:pn1} as an example. For the user served
by an MBS, SINR can be expressed as

\begin{equation}
\mathrm{SINR=\frac{\mathnormal{P_{m}h_{m,1}r_{m}^{-\alpha}}}{\mathnormal{I+\mathnormal{\sigma^{2}}}}},
\end{equation}
where $I=I_{M}+I_{s}$ and $I_{M}=\underset{i\in m}{\sum}\mathnormal{P_{m}h_{m,i}^{\mathnormal{2}}r_{m,i}^{-\alpha}}$,
$I_{s}=\underset{i\in s}{\sum}\mathnormal{P_{s}h_{s,i}^{\mathnormal{2}}r_{s,i}^{-\alpha}}$
are the interference from the MBS tier and SBS tier.

So the coverage probability when user associated with an MBS in non-cooperative
model is

\begin{eqnarray}
 &  & \mathcal{P}_{n}\left(A_{m}\right)\nonumber \\
 & = & \mathbb{P}\left(\mathrm{SINR>\theta}\right)\nonumber \\
 & = & \mathbb{\int_{\mathnormal{r>0}}P\left[\mathrm{SINR}>\theta\mid\mathnormal{r}\right]\mathnormal{f_{Rm}\left(r\right)dr}}\nonumber \\
 & \underset{=}{{\scriptstyle \left(a\right)}} & \int_{\mathnormal{r>0}}\mathbb{P}\left[h_{m,1}>P_{m}^{-1}\theta r_{m}^{\alpha}\left(I+\sigma^{2}\right)\mid\mathnormal{r}\right]f_{Rm}\left(r\right)\textrm{d}r\nonumber \\
 & = & \mathbb{\int_{\mathnormal{r>0}}E}_{I}\left[\textrm{e}^{\left(-P_{m}^{-1}\theta r_{m}^{\alpha}\left(I+\sigma^{2}\right)\right)}\right]f_{Rm}\left(r\right)\textrm{d}r\nonumber \\
 & = & \int_{\mathnormal{r>0}}\mathbb{E}\left[\textrm{e}^{\left(-P_{m}^{-1}\theta r_{m}^{\alpha}\sigma^{2}\right)}\mathcal{L}_{I}\left(P_{m}^{-1}\theta r_{m}^{\alpha}\right)\right]f_{Rm}\left(r\right)\textrm{d}r,\nonumber \\
\end{eqnarray}
where $f_{Rm}\left(r\right)$ can be seen in Lemma \ref{lem3:con fr}.
$\left(a\right)$ is because that $h_{i,j}\sim exp(1)$. $\mathcal{L}_{I}\left(s\right)$
is Laplace transform of interference $I$, it can be written as

\begin{eqnarray}
\mathcal{L}_{I}\left(s\right) & = & \mathbb{E}_{I}\left[\textrm{e}^{-sI}\right]\nonumber \\
 & = & \overset{}{\underset{i\in\left\{ s,m\right\} }{\prod}}\mathbb{E}\left[\underset{\Phi_{I_{i}}}{\prod}\textrm{e}^{-sP_{i}h_{i,j}r_{i,j}^{-\alpha}}\right]\nonumber \\
 & = & \overset{}{\underset{i\in\left\{ s,m\right\} }{\prod}}\mathbb{E}\left[\underset{\Phi_{I_{i}}}{\prod}\mathbb{E}_{h_{i,j}}\left(\textrm{e}^{-sP_{i}h_{i,j}r_{i,j}^{-\alpha}}\right)\right]\nonumber \\
 & \overset{{\scriptstyle \left(b\right)}}{=} & \overset{}{\underset{i\in\left\{ s,m\right\} }{\prod}}\mathbb{E}\left[\underset{\Phi_{I_{i}}}{\prod}\frac{1}{1+sP_{i}r_{i,j}^{-\alpha}}\right]\nonumber \\
 & \overset{{\scriptstyle \left(c\right)}}{=} & \overset{}{\underset{i\in\left\{ s,m\right\} }{\prod}}\textrm{e}^{\left(-\lambda_{i}\int_{\mathbb{R}^{2}}\left(1-\frac{1}{1+sP_{i}r^{-\alpha}}\right)\textrm{d}r\right)}\nonumber \\
 & \overset{{\scriptstyle \left(d\right)}}{=} & \overset{}{\underset{i\in\left\{ s,m\right\} }{\prod}}\textrm{e}^{\left(-2\pi\lambda_{i}\int_{d}^{\infty}\left(1-\frac{1}{1+sP_{i}r{}^{-\alpha}}\right)r\textrm{d}r\right)}\nonumber \\
 & \overset{{\scriptstyle \left(e\right)}}{=} & \overset{}{\underset{i\in\left\{ s,m\right\} }{\prod}}\textrm{e}^{\left(-2\pi\lambda_{i}\left(sP_{i}\right)^{\frac{2}{\alpha}}\right)\int_{\left(sP_{i}\right)^{-\frac{1}{\alpha}}d}^{\infty}\frac{\mu}{1+\mu^{\alpha}}\textrm{d}\mu},\nonumber \\
\end{eqnarray}
where $(b)$ uses the expression for moment generating function of
an exponential random variable, which is $h_{i,j}$; $(c)$ is due
to the probability generating functional for a PPP; $\left(d\right)$
uses the translation of surface integration. $\left(e\right)$ is
because of the variable substitution $\mu^{\alpha}=\left(sP_{i}\right)^{-1}r^{\alpha}.$
$\Phi_{I_{i}}$ is the set of all the interference BSs in tier $i$.
And the interference is expressed as the integration from $d$ to
$\infty.$ When $i\in m$, that is in MBS tier, interference is outside
the coverage area of the user's serving MBS which is the circle of
radius $r$. So we set $d=r$. As our discussion is limited in the
event of $A_{m}$, we have the condition $\mathnormal{\mathnormal{r_{s}>\left(P_{s}r^{\alpha}/P_{m}\right)^{\mathnormal{1/\alpha}}}}$.
In which, $r_{s}$ also means the distance of nearest interference
SBS. So $i\in s$, $d=r_{s}=\left(P_{s}r^{\alpha}/P_{m}\right)^{1/\alpha}.$
Similarly for $\mathcal{P}_{n}\left(B_{m}\right)$, when $i\in s$,
$\mathnormal{\overset{{\scriptstyle x_{s,k}}}{\underset{{\scriptstyle x_{s,j}\in\mathcal{C}}}{\sum}}r_{s,j}^{-\alpha}>P_{m}r_{m}^{-\alpha}/P_{s}}$,
to obtain the region of $r_{s,1}$ we have to get the approximation
$r_{s,1}^{-\alpha}>P_{m}r_{m}^{-\alpha}/P_{s}$. So $d\thickapprox\left(\frac{P_{s}}{P_{m}}\right)^{1/\alpha}$. 

\bibliographystyle{IEEEtran}
\bibliography{manu}

\begin{IEEEbiography}[{\includegraphics[height=1.25in]{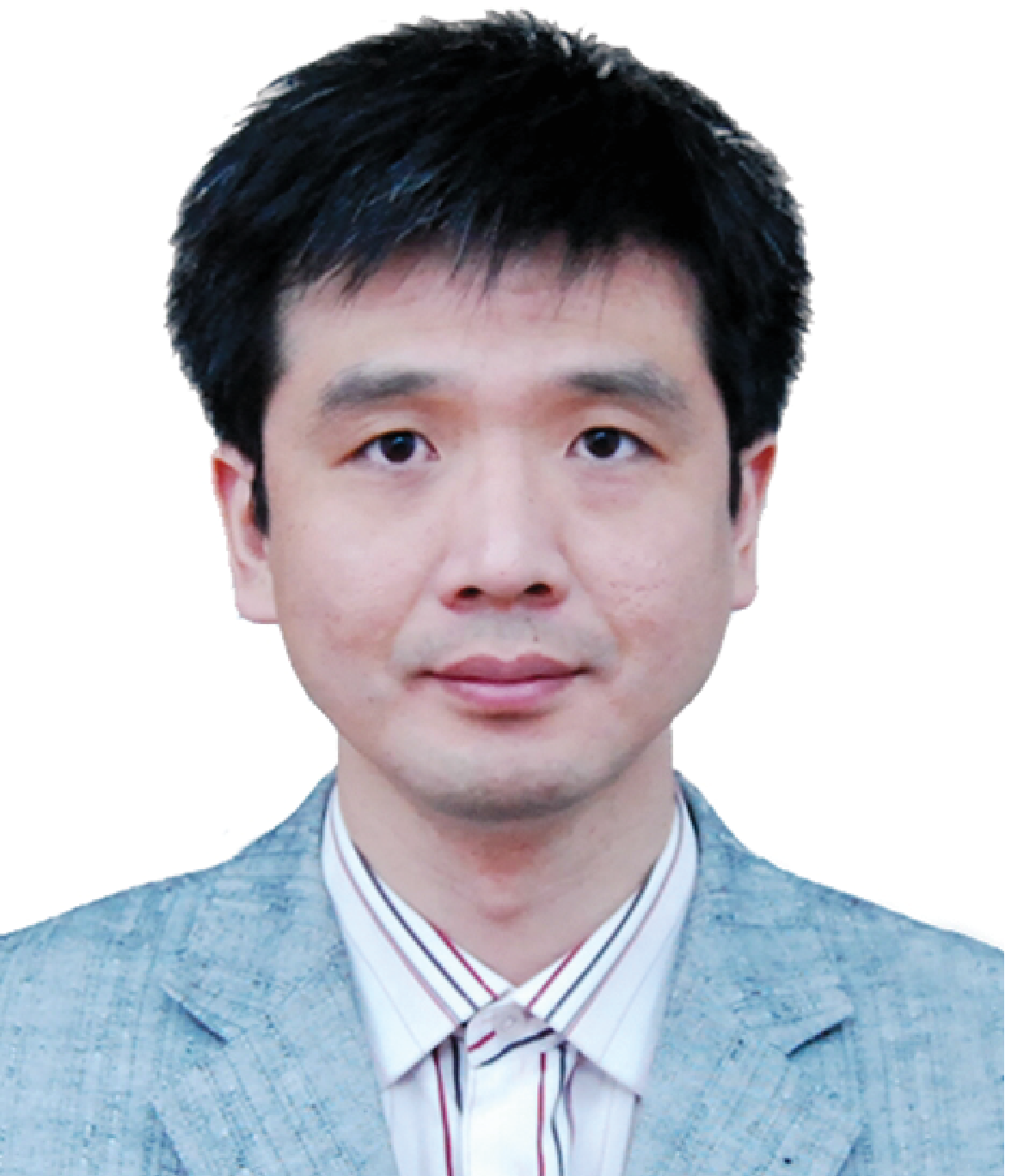}}]{Tao Han}
(M\textquoteright 13) received the Ph.D. degree in information and
communication engineering from Huazhong University of Science and
Technology (HUST), Wuhan, China, in 2001.

He is currently an Associate Professor with the School of Electronic
Information and Communications, HUST. From 2010 to 2011, he was a
Visiting Scholar with the University of Florida, Gainesville, FL,
USA, as a Courtesy Associate Professor. He has published more than
50 papers in international conferences and journals. His research
interests include wireless communications, multimedia communications,
and computer networks.

He is currently serving as an Area Editor for the \emph{European Alliance
Innovation Endorsed Transactions on Cognitive Communications}.
\end{IEEEbiography}

\begin{IEEEbiography}[{\includegraphics[height=1.25in]{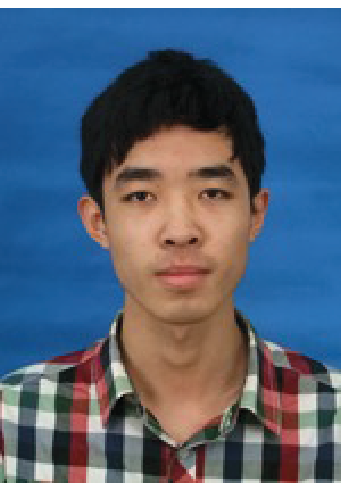}}]{Yujie Han}
received the Bachelor\textquoteright s degree in communication and
information system from Huazhong University of Science and Technology,
Wuhan, China, in 2012, where he is currently working toward the Master\textquoteright s
degree.

His research interests include cooperative communication, stochastic
geometry, and heterogeneous networks. 
\end{IEEEbiography}

\begin{IEEEbiography}[{\includegraphics[height=1.25in]{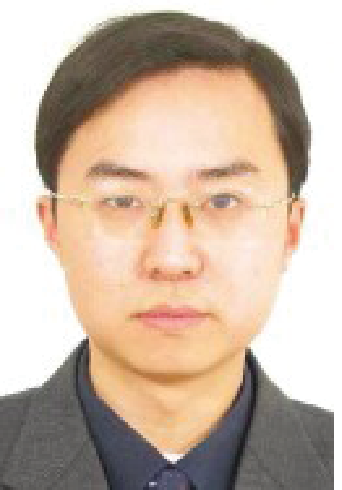}}]{Xiaohu Ge}
(M'09-SM'11) is currently a full Professor with the School of Electronic
Information and Communications at Huazhong University of Science and
Technology (HUST), China. He is an adjunct professor with the Faculty
of Engineering and Information Technology at University of Technology
Sydney (UTS), Australia. He received his PhD degree in Communication
and Information Engineering from HUST in 2003. He has worked at HUST
since Nov. 2005. Prior to that, he worked as a researcher at Ajou
University (Korea) and Politecnico Di Torino (Italy) from Jan. 2004
to Oct. 2005. His research interests are in the area of mobile communications,
traffic modeling in wireless networks, green communications, and interference
modeling in wireless communications. He has published more than 100
papers in refereed journals and conference proceedings and has been
granted about 15 patents in China. He received the Best Paper Awards
from IEEE Globecom 2010.

Dr. Ge is a Senior Member of the China Institute of Communications
and a member of the National Natural Science Foundation of China and
the Chinese Ministry of Science and Technology Peer Review College.
He has been actively involved in organizing more the ten international
conferences since 2005. He served as the general Chair for the 2015
IEEE International Conference on Green Computing and Communications
(IEEE GreenCom 2015). He serves as an Associate Editor for the \emph{IEEE
ACCESS}, \emph{Wireless Communications and Mobile Computing Journal
(Wiley)} and \emph{the International Journal of Communication Systems
(Wiley)}, etc. Moreover, he served as the guest editor for \emph{IEEE
Communications Magazine} Special Issue on 5G Wireless Communication
Systems.
\end{IEEEbiography}

\begin{IEEEbiography}[{\includegraphics[height=1.25in]{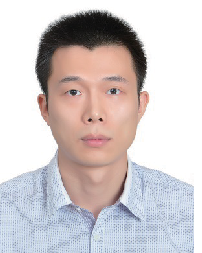}}]{Qiang Li}
(M\textquoteright 16) received the B.Eng. degree in communication
engineering from the University of Electronic Science and Technology
of China (UESTC), Chengdu, China, in 2007 and the Ph.D. degree in
electrical and electronic engineering from Nanyang Technological University
(NTU), Singapore, in 2011.

From 2011 to 2013, he was a Research Fellow with Nanyang Technological
University. Since 2013, he has been an Associate Professor with Huazhong
University of Science and Technology, Wuhan, China. He was a visiting
scholar at the University of Sheffield, Sheffield, UK from March to
June 2015. His current research interests include future broadband
wireless networks, software-defined networking, cooperative communications,
and cognitive spectrum sharing.
\end{IEEEbiography}

\begin{IEEEbiography}[{\includegraphics[height=1.25in]{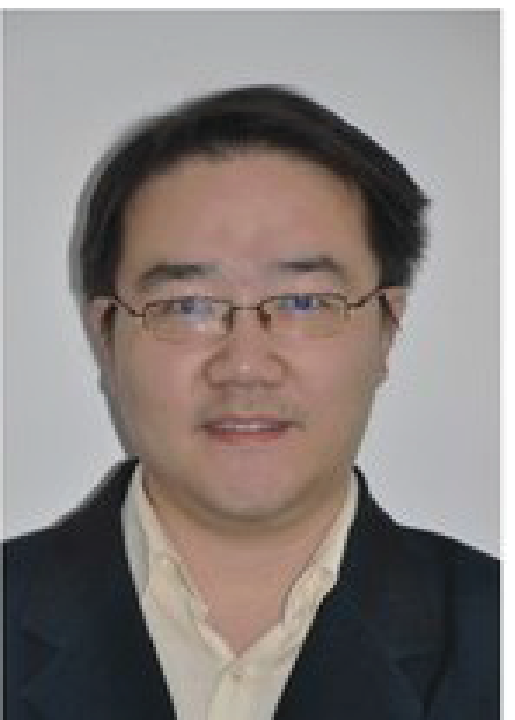}}]{Jing Zhang}
(M\textquoteright 13) received the M.S. and Ph.D. degrees in electronics
and information engineering from Huazhong University of Science and
Technology (HUST), Wuhan, China, in 2002 and 2010, respectively.

He is currently an Associate Professor with HUST. He has done research
in the areas of multiple-input multiple-output, CoMP, beamforming,
and next-generation mobile communications. His current research interests
include cellular systems, green communications, channel estimation,
and system performance analysis.
\end{IEEEbiography}

\begin{IEEEbiography}[{\includegraphics[height=1.25in]{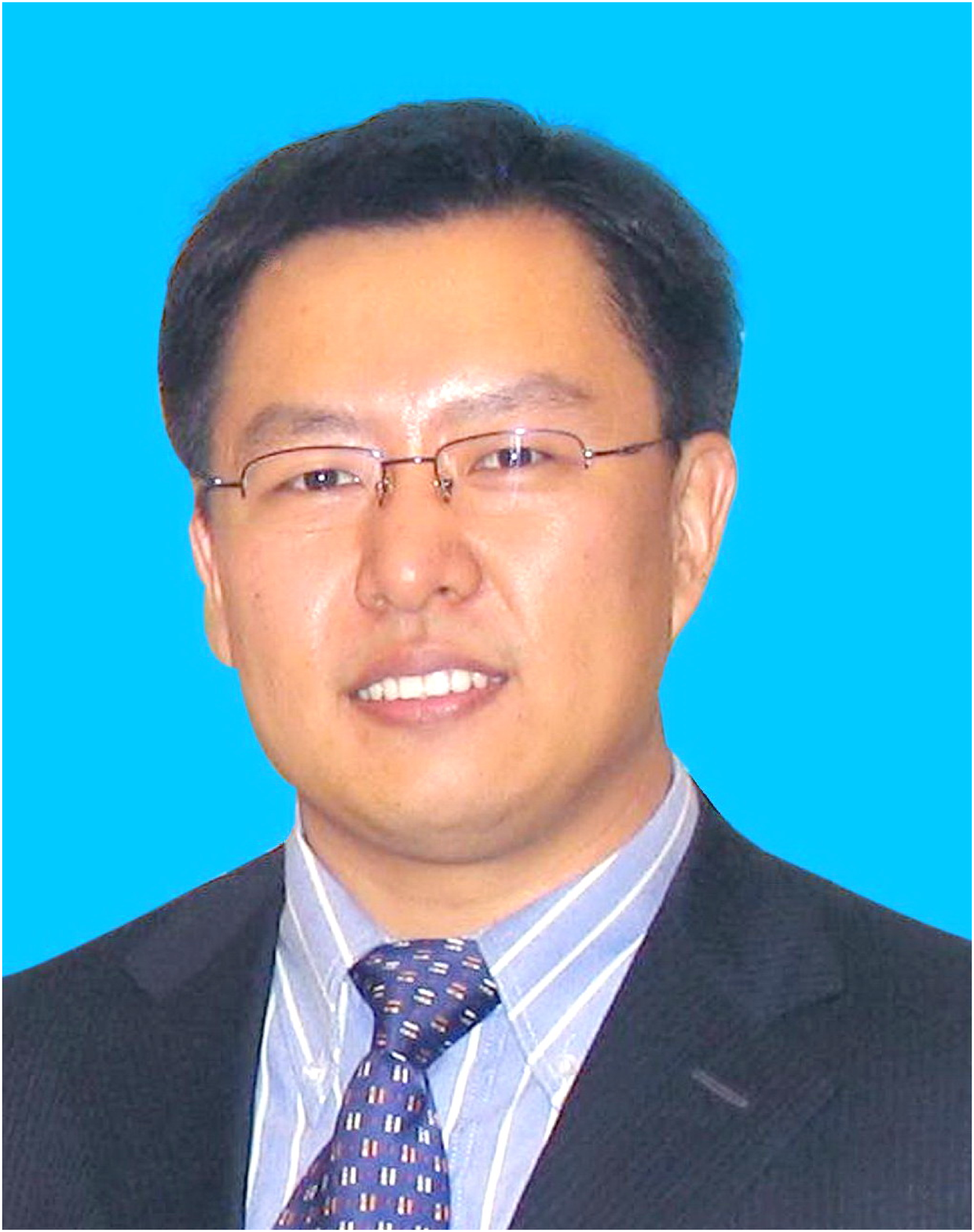}}]{Zhiquan Bai}
(M\textquoteright 08) received the M.Eng. degree in communication
and information systems from Shandong University, Jinan, China, in
2003, and the Ph.D. degree in communication engineering with honor
from INHA University in 2007, under the Grant of Korean Government
IT Scholarship, Incheon, Korea.

From 2007 to 2008, he was a postdoctor with UWB Wireless Communications
Research Center, INHA University, Incheon, Korea. Since 2007, he has
been an associate professor with the School of Information Science
and Engineering, Shandong University, China. He has published more
than 70 papers in international conferences and journals. His current
research interests include cooperative communications and MIMO system,
ultra wideband communications, cognitive radio system, and beyond-fourth
generation wireless communications.

Dr. Bai is the associate editor of the \emph{Introduction Journal
of Communication Systems}. He served as a TPC member and a session
chair for some international conferences.
\end{IEEEbiography}

\begin{IEEEbiography}[{\includegraphics[height=1.25in]{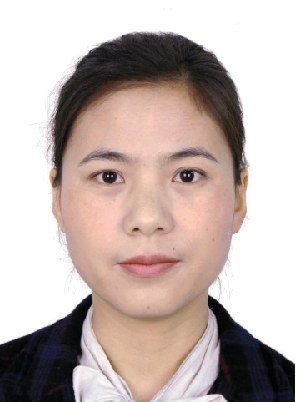}}]{Lijun Wang}
(M\textquoteright 16) received the B.S. degree in telecommunication
engineering from Xidian University, Xi\textquoteright an, China in
July, 2004 and the M.S. degree in communication and information system
from Huazhong University of Science and Technology (HUST), Wuhan,
China in June, 2008. From Sept., 2016 she will work toward the Ph.D.
degree with Wuhan University, Wuhan, China.

She is currently an associate professor with the Department of Information
Science and Technology, Wenhua College, Wuhan, China. Her research
interests include wireless communications, and multimedia communications.
\end{IEEEbiography}

\end{document}